# Spectroscopic study of interfaces in Al/Ni periodic multilayers

**Karine Le Guen, Grégory Gamblin and Philippe Jonnard**

Laboratoire de Chimie-Physique - Matière et Rayonnement, UPMC Univ Paris 06, CNRS-UMR 7614,

11 rue Pierre et Marie Curie, F-75231 Paris cedex 05, France

**Matthieu Salou, Jamal Ben Youssef, Stéphane Rioual and Bruno Rouvellou**

Laboratoire de Magnétisme de Bretagne, FRE/CNRS 2697 Université de Bretagne Occidentale,

6 Avenue Le Gorgeu, CS 93837, F-29238 Brest cedex 3, France

Using electron-induced x-ray emission spectroscopy (XES), we have studied two Al/Ni periodic multilayers that differ only by their annealing temperature: as-deposited and annealed at 115°C. Our aim is to show that XES can provide further details about the chemistry at the metal-metal interface, in addition to what is obtained by the x-ray diffraction. The distribution of valence states exhibiting Al 3p and Ni 3d character is determined from the analysis of the Al K$\beta$ and Ni L$\alpha$ emission bands respectively. The multilayer emission bands are compared to those of reference materials: pure Al and Ni metals as well as $Al_3Ni$, $Al_3Ni_2$ and AlNi intermetallics. We provide evidence that, for temperatures up to 115°C, $Al_3Ni$ is the major component of the multilayer.

**Corresponding author**:    Dr. K. Le Guen
Laboratoire de Chimie-Physique, 11 rue Pierre et Marie Curie
F-75231 Paris cedex 05, France
Tel: 33 (0)1 44 27 66 08    Fax: 33 (0)1 44 27 62 26
karine.le_guen@upmc.fr

# 1. Introduction

X-ray emission originates from the radiative decay of a core-ionized state in an atom: an electron coming from an upper level fills the core-hole and its excess energy is relaxed through the emission of a photon whose energy is equal to the energy difference between the two concerned atomic levels. Thus, the photon emission is characteristic of the emitting atom. An emission band is observed when a valence electron takes part in the electronic transition. Since valence electrons have low binding energies, the energy distribution of their emission band is very sensitive to the physico-chemical environment of the emitting atom [1-3]. Thus, the shape and width of the emission band directly reflect the chemical bonding around the emitting atom. Electron-induced x-ray emission spectroscopy (XES) is a non-destructive analysis technique which allows probing in depth the chemical composition of materials [4]. The analyzed depth depends on the penetration of the incident electrons, which is conditioned by their elastic and inelastic interactions with matter.

In the present work, we apply XES to a well-documented system in order to give evidence that this spectroscopic method allows gaining insight into the description of interface chemistry in complex materials. The Al/Ni system is an attractive model owing to the limited number of stable components, known as nickel aluminides, exhibited in the corresponding phase diagram [5]: $Al_3Ni$, $Al_3Ni_2$, $AlNi$, $Al_3Ni_5$ and $AlNi_3$. In papers devoted to the study of Al-Ni alloys formed by Ni and Al stack deposition, many parameters appear to play a predominant role such as the crystalline structure of the layers, the deposition temperature or the average atomic composition over the stack. In particular, intermixing, alloying process or formation of islands has been shown to occur upon vapor-deposition of Al on Ni [6-9] or Ni on Al [10-16]. General trends are difficult to draw from Al/Ni multilayers deposited by magnetron sputtering studies [17-22], but a high chemical reactivity is observed at the interfaces, even at room temperature. This has been recently confirmed by x-ray photoelectron spectroscopy and XES [23]. Between Ni and Al elements, Ni is known to be the main migrating element [12,19-**21, 23**]. These studies indicate that (i) multilayers have to be annealed to form stable intermetallic phases at the interfaces and (ii) even if the first aluminide to form is expected to depend on the average Al/Ni composition, the $Al_3Ni$ intermetallic compound is shown to be the first crystalline phase to form. Using grazing incidence x-ray scattering [17] or Auger electron spectroscopy (AES) depth profiling [19,21], the investigation of the structure of as-deposited Ni/Al multilayers shows the asymmetry of Ni-on-Al and Al-on-Ni interfaces, the latter being larger.

However, in spite of this high number of studies and the large variety of analytical methods used to investigate the first stage of thin films reactions in Al/Ni multilayers deposited by magnetron sputtering (X-ray reflectivity [17], AES [19-21], secondary ion mass spectroscopy [20] and Rutherford backscattering spectrometry [22]), some points remain unclear. In particular, the interpretation of AES depth profile, showing the presence of Ni in Al layer in as-deposited multilayers [19-21], is debatable since the observed effect may directly result from the ion bombardment inherent to this destructive method.

The aim of the present work is to demonstrate the complementarities of two non destructive methods, x-ray diffraction (XRD) and x-ray emission spectroscopy, to detect the presence and determine the nature of interfacial nickel aluminides in Al/Ni periodic multilayers. More precisely, we want to show that XES can provide local information about the chemistry at the metal-metal interface, concerning for instance the presence of amorphous phase. In that view, the Al/Ni multilayer system is suitable since (i) it is expected to be highly reactive and (ii) an abundant literature is available. Two samples of $Al_{0.50}Ni_{0.50}$ average composition are studied: one remains as-deposited while the other undergoes annealing at 115°C during 45 mn under vacuum. The structural changes as a result of the heating treatment are first characterized by XRD. Then, as XES allows probing into the chemical environment of the emitting atoms, the interface composition of the Al/Ni multilayers is determined from the analysis of their XES spectra. These, in turn, are fitted as a weighted sum of reference spectra (with the constraint that a minimum number of references is introduced), this methodology being now routinely used to study complex materials [23-26].

## 2. Experiment

We have studied two periodic Al/Ni multilayers, made of 10 bilayers alternating 15 nm of Al and 10 nm of Ni. They have been prepared using a calibrated RF magnetron sputtering system with high-purity Ni and Al targets in Ar gas. The base pressure was $1.10^{-7}$ mbar and the Ar working pressure was $1.10^{-2}$ mbar. The sputtering power density was 4 $W.cm^{-2}$. The multilayers were all deposited onto Si (100) substrates at room temperature. The deposition rates were estimated by thickness measurements with a mechanical profilometer. These two samples differ only by their annealing temperature: as-deposited and annealed at 115°C under vacuum during 45 mn. XRD measurements were performed using a four circles x-ray diffractometer (Brucker AXS D8) operated with the Cu Kα radiation (0.154 nm) in θ-2θ mode without channel-cut (002) Ge front monochromator. The angular resolution was 0.02°. The reference compounds used for the XES study are Al and Ni polished metals as well as $Al_3Ni$ and AlNi intermetallics. Furthermore, an Al/Ni multilayer annealed during a previous study and whose composition is known by XRD to be mainly $Al_3Ni_2$, is also used as a reference.

We have performed XES measurements using the IRIS experimental setup [27]. We have measured the Al Kβ and Ni Lα emission bands originating from the Al 3p-1s and Ni 3d-2p$_{3/2}$ transitions respectively and related to the occupied valence states having the Al 3p and Ni 3d character respectively. The energy $E_0$ of the incident electrons has been chosen to be higher than the threshold of the studied emission: in the present case 1560 and 850 eV for the Al 1s and Ni 2p$_{3/2}$ binding energies respectively. Following the ionization of the atoms present in the sample, characteristic x-rays are emitted [28,29]. They are dispersed by a high-resolution (10-10) quartz (for Al Kβ emission) or (10-10) beryl (for Ni Lα emission) bent crystal and detected in a gas-flux counter working in the Geiger regime. The spectral resolution ΔE/E is estimated to be ~ 5000 in the Al Kβ range and ~ 1500 in the Ni Lα range [27]. The current density of the electrons reaching the sample is set to less than 1 mA.cm$^{-2}$ to avoid any evolution of the sample. In order for the analysis method to remain non destructive, the current intensity has to be moderate. The shape and intensity of the studied emission are checked to remain unchanged throughout the measurements. The number of analyzed Al/Ni bilayers is estimated using the CASINO Monte-Carlo program [30], to be 10 for $E_0$ = 8 keV for the Al Kβ emission and close to 6 for $E_0$ = 5 keV for the Ni Lα emission. Given the thickness of the Al and Ni layers, the analyzed depth is ~ 250 nm for the Al Kβ emission and ~ 150 nm for the Ni Lα emission. In that latter case, we have to keep in mind that no Ni oxidation process can occur because the first Ni layer is embedded in each sample. Indeed, in each multilayer structure, Al is the top layer so that all Ni layers are kept away from air. In the following, each presented emission spectrum is normalized with respect to its maximum and a linear background, corresponding to the Bremsstrahlung contribution, is subtracted. The emission spectra of the multilayers are fitted as a weighted sum of reference spectra with the constraint that a minimum number of references is introduced.

## 3. Results and discussion

First, we analyse the composition of our two multilayer samples using XRD. The XRD patterns of the as-deposited and annealed at 115°C Al/Ni multilayers are presented in Figure 1. Before annealing (Fig. 1a), the main peaks are attributed to the elemental Al and Ni, indicating an fcc (111) texture. No other crystalline ordering, and especially no indication of the existence of an ordered Al$_3$Ni phase, can be observed. Following heat treatment at 115°C (Fig. 1b), significant structural changes are demonstrated showing that the elemental Al diffraction peak has disappeared giving way to additional weak peaks that correspond to the (220) and (102) reflections of Al$_3$Ni. No variation of the elemental Ni peak is reported.

Now we turn to the non destructive XES analysis method. It is obviously less conventional than XRD when information about interface chemistry is required, but it allows the identification of an amorphous phase or a nanometric phase for which XRD has a low sensitivity.

The Al K$\beta$ and Ni L$\alpha$ emission bands of the bulk reference compounds are presented in Figure 2. In the case of the Al K$\beta$ emission (Fig. 2a), it clearly appears that the band shape is very sensitive to the chemical environment of the Al emitting atom, as previously observed in others systems [1-3]. The shape of the emission band of Al metal is as expected in the case of a nearly free electron system: it varies approximately as the square root of the energy and exhibits a discontinuity close to 1559 eV corresponding to the Fermi level. Concerning the $Al_3Ni$, $Al_3Ni_2$ and AlNi intermetallics, the emission band is composed of two main structures: the maximum at 1556 eV is accompanied by a shoulder around 1558 eV whose intensity significantly decreases as the Al amount in the nickel aluminide decreases. In the high photon energy range, the self-absorption process reduces the intensity beyond the Fermi level. This effect is more pronounced when the Al content is higher.

In the case of the Ni L$\alpha$ emission (Fig. 2b), we have to note first that the feature at 854 eV is not directly related to the Ni 3d density of states, but to a satellite emission originating from multiply ionized states. Indeed, following the initial ionization, a Coster-Kronig process can take place leading to a multiply ionized state. Secondly, the position of the intermetallics maxima is less than 0.25 eV lower than that of Ni metal. As a consequence, these reference emission bands are not sufficiently different from each other to be used in weighted sums in order to determine if nickel aluminides are present in the studied Al/Ni multilayers. Our analysis will be essentially based on the analysis of Al K$\beta$ experimental data. However, the validity of our conclusions will be checked with the Ni L$\alpha$ results in order to check any incompatibility.

The Al K$\beta$ emission band of the two Al/Ni periodic multilayers is presented in Figure 3 where ten bilayers are probed. For the as-deposited Al/Ni multilayer, the shape of the emission band is quite similar to that of pure Al metal on the high photon energy side, but differs on the low photon energy side by the presence of an intense and wide structure centered around 1556.5 eV that is similar to the feature in the emission band of the $Al_3Ni$ intermetallic. This gives indication in the favor of the presence of $Al_3Ni$ in the Al/Ni multilayer prior to annealing. In the case of the annealed Al/Ni multilayer, the shape of the Al K$\beta$ emission band has significantly evolved. The intensity of the shoulder around 1559 eV has sharply decreased, highlighting the fact that pure Al has reacted. Using the method of the reference weighted sum that is now routine to estimate the composition of complex materials [24,25], we show in Fig. 4a that the Al Kβ emission band of the as-deposited sample is well reproduced using (44 ± 5) %wgt of $Al_3Ni$ plus (56 ± 5) %wgt of pure Al. The XRD pattern of the as-deposited multilayer shown previously (Fig. 1a) does not contain any trace of $Al_3Ni$, thus indicating that the $Al_3Ni$ phase detected by XES should be amorphous. We have to keep in mind that XES is a local probe, *i.e.* provided that the local structural environment around the Al atoms is almost the same in the amorphous and crystalline phases, their corresponding spectra will be similar. The presence of such amorphous phases has already been reported in other systems like Mo/Si multilayers [24-26]. For the sample annealed at 115°C (Fig. 4b), it appears that (i) within the uncertainty of the method (± 5%), pure Al is no more present in the multilayer meaning that the Al film has fully reacted upon annealing; (ii) the content of $Al_3Ni$ is higher than in the as-deposited sample and (iii) a small amount of $Al_3Ni_2$ has to be introduced as a component in the weighted sum. In summary, the weighted sum corresponding to the annealed sample is 92 %wgt of $Al_3Ni$ + 8 %wgt of $Al_3Ni_2$. It is not necessary to consider AlNi as a reference for the fit of the annealed sample since its shoulder at 1558 eV is too weak to reproduce the shoulder in the spectra of the annealed sample. Indeed, in the Al Kβ emission spectra presented in Figure 2(a), the intensity of the shoulder at 1558 eV decreases with the atomic fraction of Al present in the sample: 100 %at in Al, 75%at in $Al_3Ni$ and 60%at in $Al_3Ni_2$. The presence of $Al_3Ni$ as well as the fact that the whole Al amount has reacted are in good agreement with our XRD analysis (Fig. 1b). No evidence of $Al_3Ni_2$ is found in the XRD pattern. The results of the fitting procedure concerning the Al Kβ emission are summarized in Table I, where the weight fractions (% wgt) of pure metals and nickel aluminides present in each sample are given.

Despite the fact that the Ni Lα emission band of Ni, $Al_3Ni$, $Al_3Ni_2$ and AlNi are not enough characteristic to be used as references in reliable weighted sums, we have checked that an amount of at least 50 %wgt of $Al_3Ni$ has to be introduced in addition to pure Ni in order to reproduce the Ni Lα emission band of the as-deposited and annealed multilayers (not shown). The uncertainty in this case is larger than in the case of the Al Kβ emission and is estimated to be around 20 %. Here, as reported above, the Ni Lα emission band was only used to confirm the presence of the $Al_3Ni$ intermetallic within the two samples.

## 4. Conclusion

We have been able to show that the $Al_3Ni$ intermetallic is present in the Al/Ni multilayer sample, from its deposition (in a topologically disordered phase as evidenced by XES or in a nanometric or sub-nanometer phase that can be evidenced by XES but not by XRD due to its low sensitivity in this case) up to 115°C (in an ordered phase as evidenced by XRD). The aluminide easily forms even at room temperature due to the easy diffusion of the Ni atoms within the Al layers [23].

Furthermore we show by XES that the amount of $Al_3Ni_2$ in the sample increases upon annealing. In contrast, the analysis of the XRD spectra is not capable of drawing such conclusions since (i) no trace of $Al_3Ni$ is found in the as-deposited sample and (ii) no trace of $Al_3Ni_2$ is found in the annealed sample. We did not try to quantify the amount of nickel aluminides from the XRD patterns since the intensity of the corresponding peaks is weak. Furthermore, the XRD and XES quantitative results could not be related since it appears that some part of the XES signal does not come from a crystal ordered phase.

By means of x-ray emission spectroscopy, we have studied two Al/Ni multilayer samples that differ only by their annealing temperature: as-deposited and annealed at 115°C. We have identified the presence of nickel aluminides at the metal-metal interfaces in each sample. Contrary to the x-ray diffraction patterns where ordered $Al_3Ni$ is mainly present only in the annealed multilayer, the analysis of the Al Kβ emission spectra, based on linear combination of the spectra of reference compounds (pure Al metal and $Al_3Ni$, $Al_3Ni_2$ and $AlNi$ intermetallics), demonstrates the presence of $Al_3Ni$ in the as-deposited sample itself. Thus, X-ray emission spectroscopy appears to be an improved alternative to conventional methods like XRD.

**Acknowledgments**
We are grateful to B. Lépine for his help during the structural characterisation of the samples, to S.P. Pogossian for helpful discussions and to Pr. R. Gauvin from McGill University for providing us with the polished Al sample.We are grateful to B. Lépine for his help during the structural characterisation of the samples, to S.P. Pogossian for helpful discussions and to Pr. R. Gauvin from McGill University for providing us with the polished Al sample.

**REFERENCES**

[1] F. Vergand, P. Jonnard, C. Bonnelle, Europhys. Lett. 10, 67 (1998)
[2] M. Kefi, P. Jonnard, F. Vergand, C. Bonnelle, E. Gillet, J. Phys. Condens. Mat. 5, 8629 (1993)
[3] P. Jonnard, N. Capron, F. Semond, J. Massies, E. Martinez-Guerrero, H. Mariette, Eur. Phys. J. B 42, 351 (2004)
[4] P. Jonnard, J. Phys IV France 8, PR9, 33 (1998)
[5] B. Predel, O. Madelung Editor, *Phase Equilibria, Crystallographic and Thermodynamic Data of Binary Alloys*, (Springer, Berlin, 1991)
[6] D.J. O'Connor, M. Draeger and A.M. Molenbroek, Y.G. Shen, Surf. Sci. 357, 202 (1996)
[7] A. Wehner, Y. Jeliazova, R Franchy, Surf. Sci. 531, 287 (2003)
[8] S. Le Pévédic, D. Schmauss, C. Cohen, Surf. Sci. 600, 565 (2006)
[9] P. Hahn, M.F. Bertino, J.P. Toennies, M. Ritter, W. Weiss, Surf. Sci. 412, 82 (1998)
[10] L. Damoc, E. Fonda, P. Le Faure, A. Traverse, J. Appl. Phys. 92, 1862 (2002)
[11] V. Shutthanandan, A.A. Saleh, R.J. Smith, Science 450, 204 (2000)
[12] E. Fonda, F. Petroff, A. Traverse, J. Appl. Phys. 93, 5937 (2003)
[13] M.W. Ruckman, L. Jiang, M.J. Stongin, J. Vac. Sci. Technol. A 8, 134 (1990)
[14] K. Hanamoto, A. Shinya, M. Kuwahara, T. Okamoto, M. Haraguchi, M. Fukui, K. Koto, Surf. Sci. 409, 413 (1998)
[15] A. Arranz, C. Palacio, Thin Solid Films 317, 55 (1998)
[16] C. Palacio, A. Arranz, J. Phys. Chem B 104, 9647 (2000)
[17] J.D.R. Buchanan, T.P.A. Hase, B.K. Tanner, P.J. Chen, L. Gan, C.J. Powell, W.F. Egelhoff, J. Appl. Phys. 93, 8044 (2003)
[18] C. Michaelsen, G. Lucadamo, K. Barmak, J. Appl. Phys. 80, 6689 (1996)
[19] A. Zalar, S. Hofmann, D. Kohl, P. Panjan, Thin solids Films 270, 341 (1995)
[20] U. Rothhaar, H. Oechsner, M. Scheib, R. Müller, Phys. Rev. B 61, 974 (2000)
[21] A. Zalar, J. Jagielski, M. Mozetic, B. Pracek, P. Panjan, Vacuum 61, 291 (2001)
[22] E.G. Colgan, J.W. Mayer, Nucl. Instrum. Methods Phys. Res. B 17, 242 (1986)
[23] M. Salou, S. Rioual, J. Ben Youssef, D. Dekadjevi, S. Pogossian, B. Rouvellou, P. Jonnard, K. Le Guen, G. Gamblin, B. Lépine, Surf. Interface Anal. 40, 1318 (2008)
[24] I. Jarrige, P. Jonnard, N. Frantz-Rodriguez, K. Danaie, A. Bossebœuf, Surf. Interf. Anal. 34, 694 (2002)
[25] P. Jonnard, I. Jarrige, R. Benbalagh, H. Maury, J.-M. André, Z. Dankhazi, G. Rolland, Surf. Sci. 589, 164 (2005)
[26] H. Maury, P. Jonnard, J.-M. André, J. Gautier, M. Roulliay, F. Bridou, F. Delmotte, M.-F. Ravet, A. Jérome, P. Holliger, Thin Solid Films 514, 278 (2006)
[27] C. Bonnelle, F. Vergand, P. Jonnard, J.-M. André, P. Avila, P. Chargelègue, M.-F. Fontaine, D. Laporte, P. Paquier, A. Ringuenet, B. Rodriguez, Rev. Sci. Instrum. 65, 3466 (1994)
[28] L.V. Azaroff, *X-ray Spectroscopy*, McGraw-Hill Inc. (1974)
[29] C. Bonnelle, Annual Report C, The Royal Society of Chemistry, 201 (1987)
[30] P. Hovington, D. Drouin, R. Gauvin, Scanning 19, 1 (1997); Scanning, 19, 20 (1997); Scanning, 19, 29 (1997); CASINO program http://www.gel.usherbrooke.ca/casino/

**TABLE CAPTION**

Table I: Weight fractions (%wgt) of pure metals and intermetallics as determined from the fit of the Al Kβ emission band of the Al/Ni periodic multilayers. The uncertainty is estimated to be about 5 %.

**TABLE**

| Al/Ni periodic multilayer | %wgt Al | %wgt Al$_3$Ni | %wgt Al$_3$Ni$_2$ |
|---|---|---|---|
| As-deposited | 56 | 44 | - |
| T = 115°C, 45 mn | - | 92 | 8 |

Table I

**FIGURE CAPTIONS**

Figure 1: XRD pattern of the Al/Ni periodic multilayers: (a) as-deposited; (b) annealed at 115°C.

Figure 2: Al K$\beta$ (a) and Ni L$\alpha$ (b) emission bands of the reference compounds: Al and Ni pure polished metals and the AlNi, Al$_3$Ni and Al$_3$Ni$_2$ intermetallics.

Figure 3: Al K$\beta$ emission band of the as-deposited and annealed at 115°C Al/Ni periodic multilayers. Ten bilayers are probed ($E_0 = 8$ keV).

Figure 4: Comparison of the Al K$\beta$ emission band of the Al/Ni periodic multilayer with a weighted sum of reference spectra: (a) as-deposited multilayer; (b) multilayer annealed at 115°C.

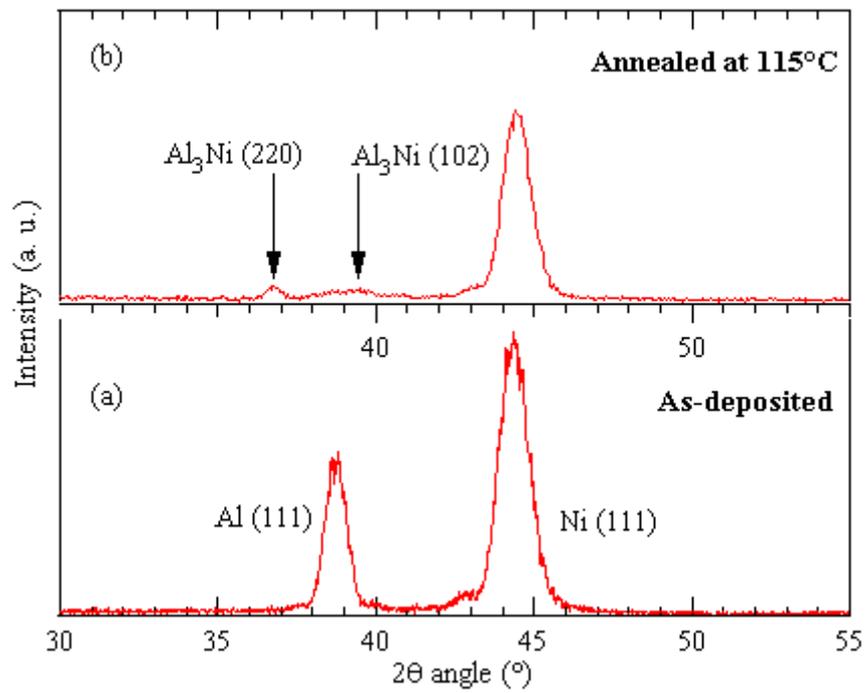

Figure 1

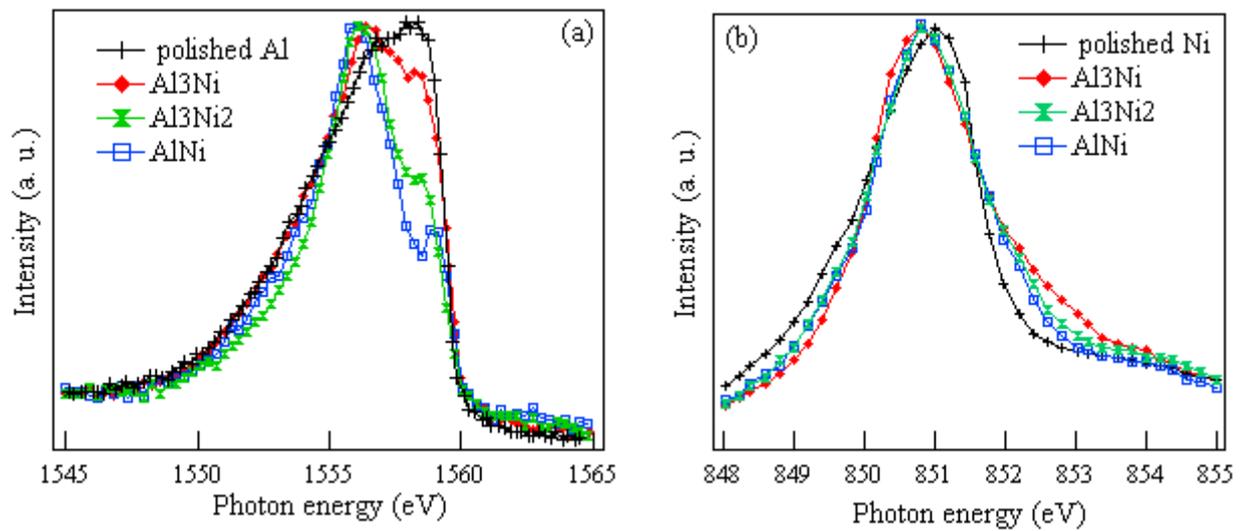

Figure 2

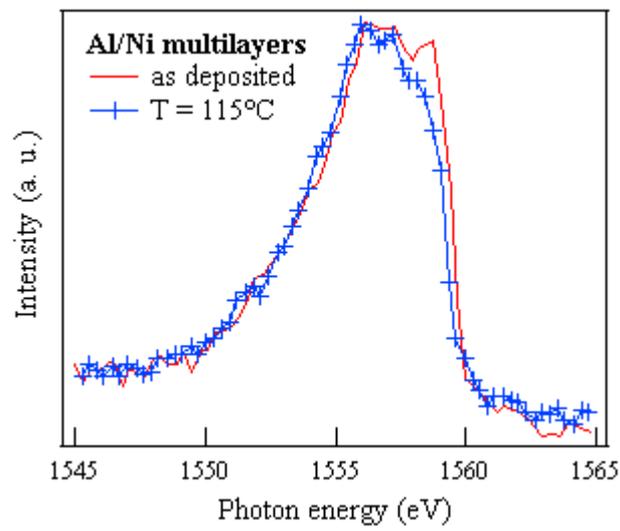

Figure 3

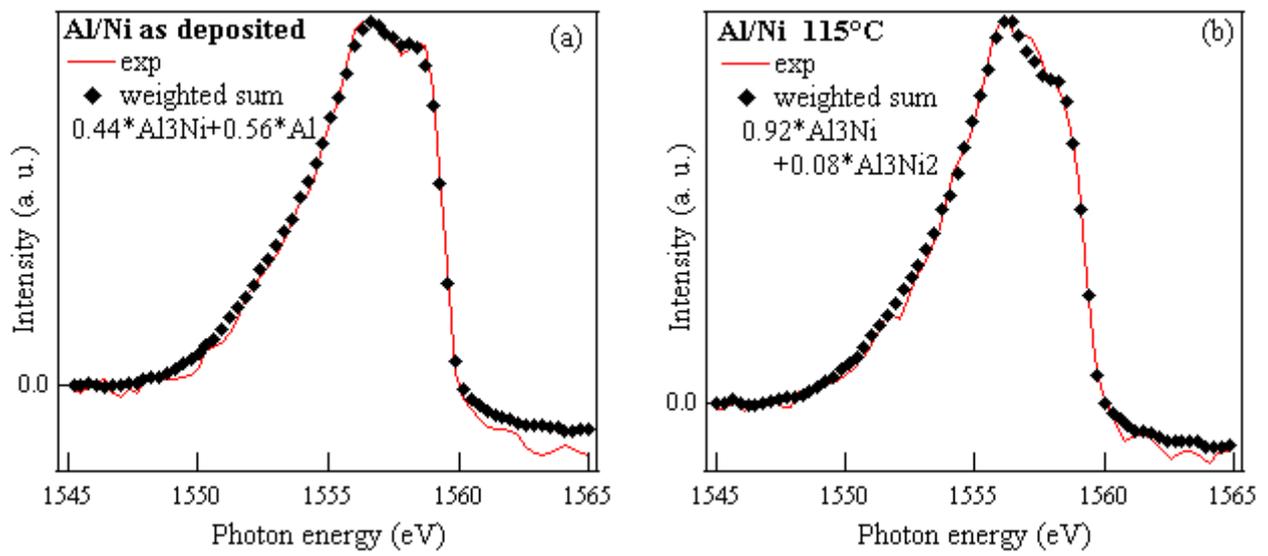

Figure 4